\documentclass[final,sort&compress]{aipproc}
\usepackage{bm}

\layoutstyle{6x9}

\begin{document}

\newcommand{\tlab}{T_{\mbox{\scriptsize lab}}}
\newcommand{\plab}{p_{\mbox{\scriptsize lab}}}
\newcommand{\pppi}{\bar p p \rightarrow \pi^-\pi^+}
\newcommand{\ppkk}{\bar p p \rightarrow K^-K^+}
\newcommand{\s}{^3\!S_1}
\newcommand{\p}{^3\!P_0}
\newcommand{\ana}{A_{0n}}
\newcommand{\dsig}{d\sigma/d\Omega}
\newcommand{\pvec}{{\mathbf p}}

\title{Lorentz contraction, geometry and range in antiproton-proton annihilation into two pions}


\author{\underline{Bruno El-Bennich}}{
  address={Laboratoire de Physique Nucl\'eaire et de Hautes \'Energies, Theory Group, \\ 
           Universit\'e Pierre et Marie Curie, Paris, France}
}

\author{W.M.~Kloet}{
  address={ Department of Physics and Astronomy, Rutgers University, Piscataway, New Jersey, USA }
}

\begin{abstract}

We present a geometric interpretation of the so-called annihilation range in reactions of the type 
$\bar pp \rightarrow$ {\em two light mesons\/} based upon Lorentz effects in the highly relativistic 
final states ($\gamma=E_{\mathrm{cm}}/2mc^2\simeq 6.8-8.0$). Lorentz-boosted meson wave functions, 
within the framework of the constituent quark model, result in a richer angular dependence of the 
annihilation amplitudes and thus in higher partial wave contributions ($J>1$) than usually obtained.  
This approach sheds some light on what could be a "{\em short}" annihilation range and how it is
influenced by the angular distribution of the final states.
\end{abstract}

\maketitle



In this talk we summarize the results obtained in Refs.~\cite{paper1,paper2,paper3} while focusing 
on the details of Lorentz effects in reactions of the type $\pppi$ or $\ppkk$ when the final states
are highly energetic. The initial motivation is twofold: for one, a long-standing debate about the
antiproton-proton range exists in which proponents of a {\em very\/} short annihilation range, of the
order of the Compton wavelength of the annihilating baryons ($r_{\mathrm{ann.}}=m_N^{-1} \simeq 0.1$~fm), 
emphasize the role of analytical properties of Feynman graphs~\cite{shapiro}. This should be {\em a priori\/}
the range of baryon exchange in the $t$-channel.

Comparison of baryon exchange model results \cite{moussalam,mull1,mull2,Yan1} with $\pppi$ and 
$\ppkk$ data on differential cross sections $\dsig$ and analyzing powers $\ana$, as measured in the 
pre-LEAR experiments at CERN and KEK~\cite{Eisenhandler,Carter,Tanimori} as well as at LEAR~\cite{hasan1},
nonetheless clearly indicates that these models are too short ranged. They give vanishing total 
$\bar pp$ angular momentum  $J\geq 1$ contributions to the annihilation amplitudes, whereas partial wave
analyzes \cite{Oakden,hasan2,Kloet1,Martin1} of both the pre-LEAR~\cite{Eisenhandler,Carter,Tanimori} 
and LEAR~\cite{hasan1} data consistently point to higher partial wave amplitudes with $J$ up to 3 or 4. 
Of course, compared with elastic proton-proton scattering, where partial waves up to $J=9-10$ are necessary 
to fit the data, antiproton-proton annihilation is a short-range interaction. Yet, the experiments 
mentioned above seem not to confirm the {\em extreme\/} shortness implied by the baryon exchange models. 
It is worthwhile to note that the annihilation range is not merely given by the Compton wavelength of the 
baryons alone. In general, the annihilation vertices are regularized with appropriate form factors, which 
smear out the baryon-meson interaction. For example, in the case of Ref.~\cite{Yan1} the form factor of the
pseudovector pion-proton coupling is found to improve the fits to $\pppi$ cross sections and analyzing 
powers. It is therefore not advisable to ascribe the annihilation range solely to heavy baryon exchange 
in the $t$-channel.

The use of alternative geometric arguments, such as the minimal overlap for annihilation of an $\bar pp$ 
pair within a quark picture motivated by bag models, was investigated by Alberg \textit{et al.\/}~\cite{alberg}. 
In their fit to $\bar pp$ scattering data the effective nucleon bag radius $R$ takes on values of 
$0.6-1.1$~fm. The crucial assumption is that $\bar pp$ annihilation proceeds only when the $\bar p$ and 
$p$ bags overlap so the quarks and antiquarks can rearrange or annihilate to form the observed mesons. 
In Ref.~\cite{alberg} the annihilation is found to peak at $0.7$~fm or $0.95$~fm depending on the bag-model 
form used. Hence, in geometric terms the annihilation range is roughly twice the nucleon radius which 
is much larger than the Compton wavelength of the nucleon. If one uses a chiral bag model then the pressure 
of the mesonic cloud about the core quarks decreases the nucleon bag radius. Since the external mesons are 
not believed to participate in the $\bar pp$ annihilation, the upper bound for the annihilation range is 
somewhat lowered. Nonetheless, quark model calculations also fail to reproduce the LEAR data \cite{Kloet2,muhm96} 
as they too are rather short-ranged. The range, in this case, is an intrinsic property of overlap integrals 
involving (anti)quark wavefunctions with size parameters that reproduce the particle radii known from fits
as the one just mentioned~\cite{alberg}.

Yet another approach to define an annihilation range is due to Povh and Walcher~\cite{povh} and was 
subsequently applied by the Bonn group~\cite{haidenbauer} to the Bonn and Paris $\bar NN$ potentials. 
In a nutshell, the method consists of taking the quantity $p_l(r)=2/\hbar~W_{\bar NN} R_l^*(r)R_l(r)r^2$ as
the annihilation probability for a partial wave of angular momentum $l$, where $R_l(r)$ is the radial
component of the (Bonn or Paris) $\bar NN$ wavefunction and $W_{\bar NN}$ the imaginary part of the potential. 
Plotting $p_l(r)$ for several partial waves, they found for the Paris $\bar NN$ potential a maximum localized 
at about 0.5~fm and for the Bonn $\bar NN$ potential this value is about 1.1~fm. This range compares well 
with the quark model prediction discussed in the previous paragraph. 

The second motivation stems from the observation that the Hasan {\em et al.\/} LEAR data~\cite{hasan1} reveals 
a strong left-right asymmetry with respect to the beam direction and displays large variations of the cross 
sections as a function of the c.m. angle. In our work, we try to determine how relativistic effects, namely 
Lorentz contractions in quark-model wavefunctions, influence the annihilation amplitudes and whether these 
effects are relevant to the annihilation range.\vspace*{-3mm}

\section{Quark Model and intrinsic meson wavefunctions}

In a quark model description of the $\bar pp$ annihilation process, one usually seeks guidance from Feynman 
diagrams in order to deduce appropriate transition operators ({\em i.e.\/} the operators linking the initial
$\bar pp$ quarks and antiquarks so they form the right $\bar qq$ pairs). These so-called quark-line diagrams
(QLD) are classified according to their flavor-flux topology into rearrangement or annihilation diagrams. 
In the former case a $\bar qq$ pair is annihilated and a quark and antiquark are rearranged to produce the
final mesons, while in the latter case two $\bar qq$ pairs are annihilated and an $\bar qq$ state is created 
in the final state. The $\bar qq$ pairs annihilate into states with distinct quantum numbers $J^P$ and $I$. 
In a way, one can conceive of the model as an expansion of the annihilation amplitude in terms of increasing 
total angular momentum $J^P$. Parity requires the $\bar qq$ pairs be annihilated/created in an $S=1$ state. 
Thus the spin-multiplicity is fixed and for $J^P=0^+,1^-...$ one gets $^3P_0$, $^3S_1$... annihilation operators.
For a concise review of the QLD see, for instance, Ref.~\cite{dover}. The resulting transition operators 
may be written in a Hamiltonian form for both the $\p$ and  $\s$ cases as
{\setlength\arraycolsep{2pt}
\begin{eqnarray} 
\mathcal{H}(\p) &=& \gamma \sum_{ijmn} a_m^\dagger(\mathbf{k}') a_n (\mathbf{k}) a_i (\pvec) b_j (\pvec')\, 
    \bm{\sigma}\!\cdot\!(\pvec-\pvec')\, (2\pi)^3 \delta (\mathbf{k}'-\mathbf{k}-\pvec'-\pvec) + \mathrm{h.c.}
\hspace*{8mm} \\
\mathcal{H}(\s) &=& \kappa \sum_{ijmn} a_m^\dagger(\mathbf{k}') a_n (\mathbf{k}) a_i (\pvec) b_j (\pvec')\, 
 \Big [ -2\, \bm{\sigma}_{ij}\!\cdot\!\mathbf{k} +i\,(\bm{\sigma}_{mn}\times\bm{\sigma}_{ij}) 
 \!\cdot\!(\mathbf{k}-\mathbf{k}')
 \Big ]\,\times  \nonumber\\ 
 & & \times\; (2\pi)^3 \delta (\mathbf{k}'-\mathbf{k}-\pvec'-\pvec) + \mathrm{h.c.} ,
\end{eqnarray}}\vspace*{-5mm}

\noindent
where color indices have been suppressed and $ijmn$ sums over flavor states. The delta-function imposes 
momentum transfer from the annihilated $\bar qq$ pair to one of the remaining (anti)quarks. Additionally, 
color is exchanged in the $\s$ mechanism since $J^P=1^-$ is the quantum number of the gluon. The relative 
strength $\lambda=|\gamma/\kappa|$ is a fit parameter.

Confinement is simulated in a harmonic oscillator basis. The single (anti)quark wavefunctions are therefore
bundled in Gaussian wave packets in which the c.m. motion of the (anti)proton and pions can be separated from the 
relative motion of their respective quarks and antiquarks. The total wavefunction of the pion, for example, 
writes in momentum space as
\begin{eqnarray}
 \phi_\pi(\pvec_{i},\pvec_{j}) & = & N_\pi \left ( \frac{4\pi}{\beta}\right )^{\!\!3/2}\!\! u(\pvec_i)v(\pvec_j)
 \chi_{\mathrm{iso}}\, \chi_{\mathrm{color}}\; \exp \Big \{\! -\frac{1}{2\beta}\!\sum_{\mu={i, j}}  
 \Big [ \pvec_\mu -\mbox{$\frac{1}{2}$}\mathbf{Q_\pi}\Big ]^2 \Big \}. \hspace*{4mm}
\label{pion}
\end{eqnarray} 
Here, $\pvec_i$ and $\pvec_j$ are the $\bar qq$ momenta and $\mathbf{Q_\pi}=\pvec_i+\pvec_j$ is the pion 
momentum in the c.m. frame whereas $u(\pvec_i)$ and $v(\pvec_j)$ are the usual quark and antiquark Dirac spinors. 
The size parameter $\beta$ is chosen so as to give the pion a radius $r_\pi=0.48$~fm.

\section{Lorentz Effects in Final States}

So far, we have dealt with the wave representation of the pion in its rest frame where it is spherical (meson in an 
$s$-wave). In $\pppi$, on the other hand, the pions are produced at considerable c.m. energies already at very low 
$\bar p$-beam energies ( $\sqrt{s}=1.9-2.2$~GeV in Ref.~\cite{hasan1}). The large mass difference between the pion 
and the proton causes the final state to be highly relativistic, as can be simply seen from the relativistic factor 
$\gamma=\sqrt{s}/2m_\pi c^2\simeq 6.8-8.0$. In computing the annihilation amplitudes (and observables), one puts 
oneself in the c.m. frame in order to compare with experiment. Therefore, {\em all} ingredients must be transformed 
to the c.m. For the pions, this means they are not spherical anymore, since they are boosted from their rest frame 
to the c.m. frame. Eq.~(\ref{pion}) cannot be used as such and instead the general Lorentz transformation 
$\pvec_i'\equiv l^{-1}\pvec_i$ (where $\pvec_i$ is a quark momentum in the pion rest frame) yields
\begin{eqnarray}
\phi_\pi(l^{-1} \pvec_{i},l^{-1}\pvec_{j})
 & = & \frac{N_\pi}{\sqrt{\gamma}} \Big(\frac{4\pi}{\beta}\Big)^{\!\!3/2}\!\!
 d^{1/2}_{\lambda\lambda'}(\theta) u(l^{-1}\pvec_i,\lambda) d^{1/2}_{\lambda\lambda'}(-\theta) v(l^{-1} \pvec_j,\lambda)
 \chi_{\mathrm{iso}}\, \chi_{\mathrm{color}}\times\nonumber \\
 & & \times\; \exp \Big \{ -\frac{1}{4\beta} \Big [ (\pvec_i-\pvec_j)^2 + \left (\frac{1}{\gamma^2}-1\right ) \!
   \left ( [\pvec_i-\pvec_j]\!\cdot\mathbf{\hat Q_\pi}\right )^{\!2}\Big ]\Big\} .
\label{expboost}
\end{eqnarray}
In the exponent, the momentum components along the boost direction $\mathbf{\hat Q_\pi}=\mathbf{Q_\pi}/|\mathbf{Q_\pi}|$
have been separated from the perpendicular ones. Naturally, for $\gamma=1$ one retrieves the original wavefunction 
in Eq.~(\ref{pion}). The additional factor $1/\sqrt{\gamma}$ is due to proper normalization to one of the 
wavefunction. The Lorentz transformations are effected on the Dirac spinors via the Wick rotations 
$d^{1/2}_{\lambda\lambda'}(\theta)$ in spin space, where the Wick angle $\theta$ depends on the relativistic factor 
$\beta=v/c$ rather than on $\gamma$. The indices $\lambda$ and $\lambda'$ are the (anti)quark helicities in the pion 
rest frame and in the c.m. frame, respectively. We have applied the c.m. equal-time condition $t_i'=t_j'$ for the time 
components of the quark and antiquark, which is also frequently used in Bethe-Salpeter approaches. The effect of the 
time components is currently under investigation. While the boosting of spinors gives rise to subtle interference 
effects in the annihilation amplitudes (see discussion in Ref.~\cite{paper2}), its magnitude is negligible compared 
with the $\gamma$-components. Once the wavefunctions of Eq.~(\ref{expboost}) are employed to derive the annihilation 
amplitudes, it becomes clear that the boost effects are dramatic. In summary, the annihilation amplitudes, being of 
Gaussian form, acquire new relativistic terms. Their magnitude can be, depending on the $\gamma$ (and hence $\sqrt{s}$) 
value, up to two orders of magnitude larger than the one of non-relativistic terms. Furthermore, this strong 
$\gamma$-dependence is not constant. The boosts also introduce an angular dependence by means of the projection on 
the boost direction $\mathbf{\hat Q_\pi}$ in the exponential. It is this dependence (which at the hadronic level 
equates to the c.m. angle between the antiproton beam and an outgoing pion) that is novel and crucial to any 
improvement in a fit to the LEAR data.

\begin{figure}[t]
\includegraphics[scale=0.53]{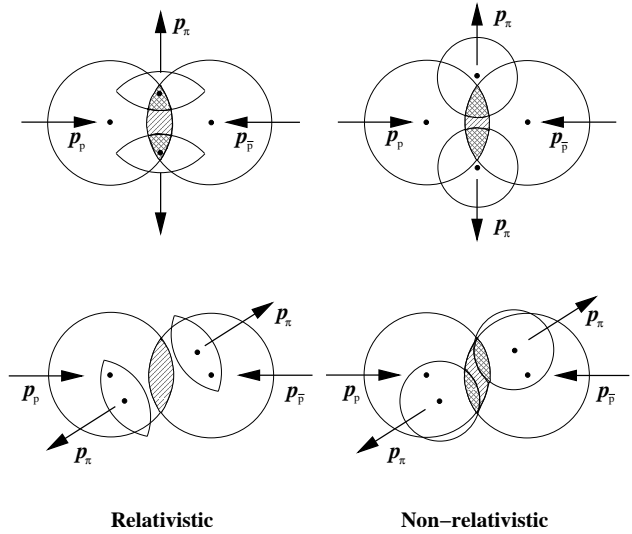}
\caption{Geometry of the overlap in the $\pppi$ annihilation. See text for details.}
\label{fig1}
\end{figure}

\section{Annihilation and Geometry}

In the following, we propose a somewhat different geometric interpretation. The strong modification of both 
the $\p$ and $\s$ annihilation amplitudes is entirely due to the Lorentz contractions in the pions. In the
overlap integrals this implies a narrowed overlap of the initial $\bar pp$ pair and the final pion wave functions. 
At first sight, this looks discouraging since our introductory discussion revolved around how to increase the 
annihilation range in a quark model. Yet, this decrease in overlap is merely one piece of the puzzle. 
The other part is illustrated in Fig.~\ref{fig1} where tentatively the angle dependence of the 
amplitudes due to relativistic effects is shown for two cases. The smaller blobs represent the pions and are 
round for $\gamma=1$ and oval (Lorentz contracted) otherwise. The distances between the two pions and the proton 
and antiproton are the same in each picture. However, in the first case the pions are produced perpendicularly 
to the $\bar p$ beam direction and all particles overlap. They do so somewhat less when the pions are contracted. 
In the second case, where the pions are emitted at a different angle, the four spheres overlap if relativity is 
neglected while the two pion-ellipsoids overlap much less with the the $\bar pp$ spheres if relativity is included.
This angle-dependent overlap is responsible for a richer angular dependence of the annihilation
amplitudes. We therefore obtain significant contributions to $J\geq 1$ partial waves not present previously
and a fit to the LEAR data~\cite{hasan1} is quite successful~\cite{paper3}.
On the other hand, it is clear that one cannot define an annihilation range \textit{per se} since this range
depends on the pion direction with respect to the initial $\bar p$-beam. Even so, one can conclude that in order
for the $\bar pp$ pair to annihilate, the quarks and antiquarks must have a considerable overlap whose size is 
less than two nucleon radii.


\begin{theacknowledgments}
We appreciated helpful discussions with Mary Alberg, Thomas Gutsche, Beno\^{\i}t Loiseau, Johann Haidenbauer, 
Fred Myhrer and Slawomir Wycech.
\end{theacknowledgments}



\bibliographystyle{aipproc}   


\end{document}